%% file: RuCl3_combined.tex
\documentclass[twocolumn,prl,showpacs,superscriptaddress,preprintnumbers,amsmath,amssymb,floatfix]{revtex4}
\usepackage{graphicx}
\usepackage{dcolumn}
\usepackage{bm}
\usepackage{color}
\usepackage{ragged2e}
\usepackage{subfiles}
\begin{document}

\subfile{RuCl3_main.tex}

\subfile{RuCl3_supp.tex}

\end{document}

%% file: RuCl3_main.tex
\title{Electronically highly cubic conditions for Ru in $\alpha$-RuCl$_{3}$}
\author{S.~Agrestini}
  \affiliation{Max Planck Institute for Chemical Physics of Solids,
     N\"othnitzerstr. 40, 01187 Dresden, Germany}
\author{C.-Y.~Kuo}
  \affiliation{Max Planck Institute for Chemical Physics of Solids,
     N\"othnitzerstr. 40, 01187 Dresden, Germany}
\author{K.-T.~Ko}
  \affiliation{Max Planck Institute for Chemical Physics of Solids,
     N\"othnitzerstr. 40, 01187 Dresden, Germany}
\author{Z.~Hu}
  \affiliation{Max Planck Institute for Chemical Physics of Solids,
     N\"othnitzerstr. 40, 01187 Dresden, Germany}
\author{D.~Kasinathan}
  \affiliation{Max Planck Institute for Chemical Physics of Solids,
     N\"othnitzerstr. 40, 01187 Dresden, Germany}
\author{H.~Babu~Vasili}
  \affiliation{ALBA Synchrotron Light Source, E-08290 Cerdanyola del Vall\`{e}s, Barcelona, Spain}
\author{J.~Herrero-Martin}
  \affiliation{ALBA Synchrotron Light Source, E-08290 Cerdanyola del Vall\`{e}s, Barcelona, Spain}
\author{S.~M.~Valvidares}
  \affiliation{ALBA Synchrotron Light Source, E-08290 Cerdanyola del Vall\`{e}s, Barcelona, Spain}
\author{E.~Pellegrin}
  \affiliation{ALBA Synchrotron Light Source, E-08290 Cerdanyola del Vall\`{e}s, Barcelona, Spain}
\author{L.-Y.~Jang}
  \affiliation{National Synchrotron Radiation Research Center, 101 Hsin-Ann Road, Hsinchu 30076, Taiwan}
\author{A. Henschel}
 \affiliation{Max Planck Institute for Chemical Physics of Solids,
     N\"othnitzerstr. 40, 01187 Dresden, Germany}
\author{M. Schmidt}
 \affiliation{Max Planck Institute for Chemical Physics of Solids,
     N\"othnitzerstr. 40, 01187 Dresden, Germany}
\author{A.~Tanaka}
  \affiliation{Department of Quantum Matter, ADSM, Hiroshima University, Higashi-Hiroshima 739-8530, Japan}
\author{L.~H.~Tjeng}
  \affiliation{Max Planck Institute for Chemical Physics of Solids,
     N\"othnitzerstr. 40, 01187 Dresden, Germany}

\date{\today}
\begin{abstract}
We studied the local Ru $4d$ electronic structure of $\alpha$-RuCl$_3$ by means of polarization
dependent x-ray absorption spectroscopy at the Ru-$L_{2,3}$ edges. We observed a vanishingly
small linear dichroism indicating that electronically the Ru $4d$ local symmetry is highly cubic.
Using full multiplet cluster calculations we were able to reproduce the spectra excellently and
to extract that the trigonal splitting of the $t_{2g}$ orbitals is $-12\pm10$~meV, i.e. negligible
as compared to the Ru $4d$ spin-orbit coupling constant. Consistent with our magnetic circular dichroism
measurements, we found that the ratio of the orbital and spin moments is 2.0, the value expected
for a $J_{\emph{eff}}=1/2$ ground state. We have thus shown that as far as the Ru $4d$ local properties
are concerned, $\alpha$-RuCl$_3$ is an ideal candidate for the realization of Kitaev physics.
\end{abstract}


\maketitle

Geometrically frustrated quantum spin systems are important owing to the fact that frustration
often results in a suppression of conventional mean field ground states in favor of more exotic
phases of matter. Current research focuses on the effect of spin-orbit coupling (SOC) and the
role it plays in the realization of different exotic phases such as unconventional superconductivity
or quantum spin liquids \cite{Singh12,Chaloupka10,Reuther11}. Especially, quantum spin liquids
can result in topological states with fractional excitations. An important, theoretically solvable
model is the Kitaev model with spin-1/2 on a honeycomb lattice, where the coupling between
neighboring spins is highly anisotropic with bond-dependent spin interactions. In contrast to
spin liquids arising from usual geometrical frustrated spin arrangements, the bond-dependent
spin interactions within the Kitaev model frustrate the spin configuration on a single site
\cite{Kitaev06}.

The search for fractionalized excitations and the identification of a Kitaev spin liquid state has
been experimentally quite difficult. Increased attention has been
focussed on the honeycomb iridates \cite{Singh2010,Gretarsson2013}, starting from the
assumption that large spin-orbit coupling is the leading energy scale in determining the ground
state such that the Ir 5$d$ $t_{2g}$ orbitals are described in terms of $J_{\emph{eff}}=1/2$ and
3/2 orbitals. However, the real iridate systems exhibit trigonal distortion
($D_{trig}= 0.1$~eV \cite{Gretarsson2013}) and a significant itinerant character of the Ir $5d$
orbitals \cite{Mazin2012,Foyevtsova2013,Agrestini2017}, which complicates the electronic
ground state.
Despite a flurry of both theoretical and experimental studies,
the nature of the ground state in honeycomb iridates are being fiercely debated
and the occurrence of Kitaev physics is still far from clear.

Recently, $\alpha$-RuCl$_3$ has been suggested as a promising candidate material for the
realization of the Kitaev model \cite{Plumb14} and excitations observed via
Raman \cite{Sandilands15,Nasu16} and inelastic neutron scattering \cite{Banerjee15} have
been presented as evidence that $\alpha$-RuCl$_3$ may be close to a quantum spin liquid
ground state.  In the last two years a number of publications discussing the realization of the Kitaev physics in $\alpha$-RuCl$_3$ has appeared in literature \cite{Johnson15,Kubota15,Majumder15,HSKim15,Rousochatzakis2015,Cao16,Yadav16,Sandilands16,Winter2016,Chaloupka2016,HSKim16,Lang2016,Koitzsch2016,Janssen2016,Sizyuk2016,Zhou2016,Ziatdinov2016,Sears2016,Sinn2016,Ran2016,Catuneanu2016,Sandilands2016b}. $\alpha$-RuCl$_3$ has a monoclinic structure, where the Ru atoms are arranged
in nearly regular honeycomb planes with a Ru-Cl-Ru bond close to 90$^{\circ}$, the latter
being one of the conditions for the realization of Kitaev magnetism. The Ru$^{3+}$ ions in
$\alpha$-RuCl$_3$ (hereafter RuCl$_3$) have the same $t_{2g}^5$
configuration as Ir$^{4+}$ ions in the iridates. The SOC, despite being modest ($\sim$150~meV),
is still thought to be the leading energy scale and able to generate a $J_{\emph{eff}}=1/2$ ground state.

Unfortunately, a precise determination of the atomic positions of the Cl ions by x-ray diffraction (XRD) is
quite difficult with conflicting reports about the crystallographic
structure of RuCl$_{3}$ \cite{Stroganov1957,Banerjee15,Johnson15,Cao16}
in literature due to the broad mosaicity arising from the weak
Van-der-Walls bond existing between the layers. The intensity of Bragg peaks in XRD is strongly
affected by the diffuse scattering produced by twins and sliding stacking faults. We will refer
in the following to the last diffraction study \cite{Cao16}, which was performed on untwinned
RuCl$_3$ single crystals with moderate stacking faults. According to this investigation, the
local structure is close to cubic despite the low symmetry of the point group of the Ru site,
and the dominant distortion of the RuCl$_3$ octahedra is trigonal with the trigonal axis normal
to the $ab$ plane. Additional tetragonal distortions are present but negligible \cite{Cao16}.

Notwithstanding the moderate trigonal distortion, quantum chemistry calculations using
the structure given by Ref. \onlinecite{Cao16} proposed a complete lifting of the degeneracy
of the $t_{2g}$ orbitals by a trigonal splitting of $D_{trig}=70$~meV \cite{Yadav16}.
Experimentally, a splitting of the order of the SOC was estimated from the large anisotropy shown
by high field magnetization measurements \cite{Kubota15}. Raman scattering spectroscopy
observed a single peak instead of the two-peak structure characteristic for trigonal distortion,
which might indicate a nearly cubic local symmetry but could also be explained with the
zero-intensity of one peak for symmetry reasons (e.g. selection rules) \cite{Sandilands16}.
Considering the critical importance of the local symmetry for the realization of the Kitaev physics,
there is a clear need to establish in a quantitative way the magnitude of the trigonal distortion
and its effect on the magnetic ground state of RuCl$_{3}$. Theoretical studies in the literature
have shown that the analysis of the ground state of RuCl$_{3}$ heavily relies on the trigonal
field strength relative to the SOC \cite{Winter2016,Chaloupka2016}.

Here, we report on a Ru $L_{2,3}$ edge x-ray absorption spectroscopy (XAS) study of the local
electronic and magnetic state of the Ru$^{3+}$ ion in RuCl$_{3}$, using both linear and
circular polarized light. In combination with full-multiplet cluster simulations using parameters
which are based on $\emph{ab-initio}$ band structure calculations, we can extract values for the
trigonal crystal field splitting as well as the ratio between the orbital and spin contributions
to the local magnetic moment, thereby evaluating to what extent the local $J_{\emph{eff}}=1/2$
ground state is realized for the Ru ion.

\begin{figure}[b]\centering
\includegraphics[width=\linewidth]{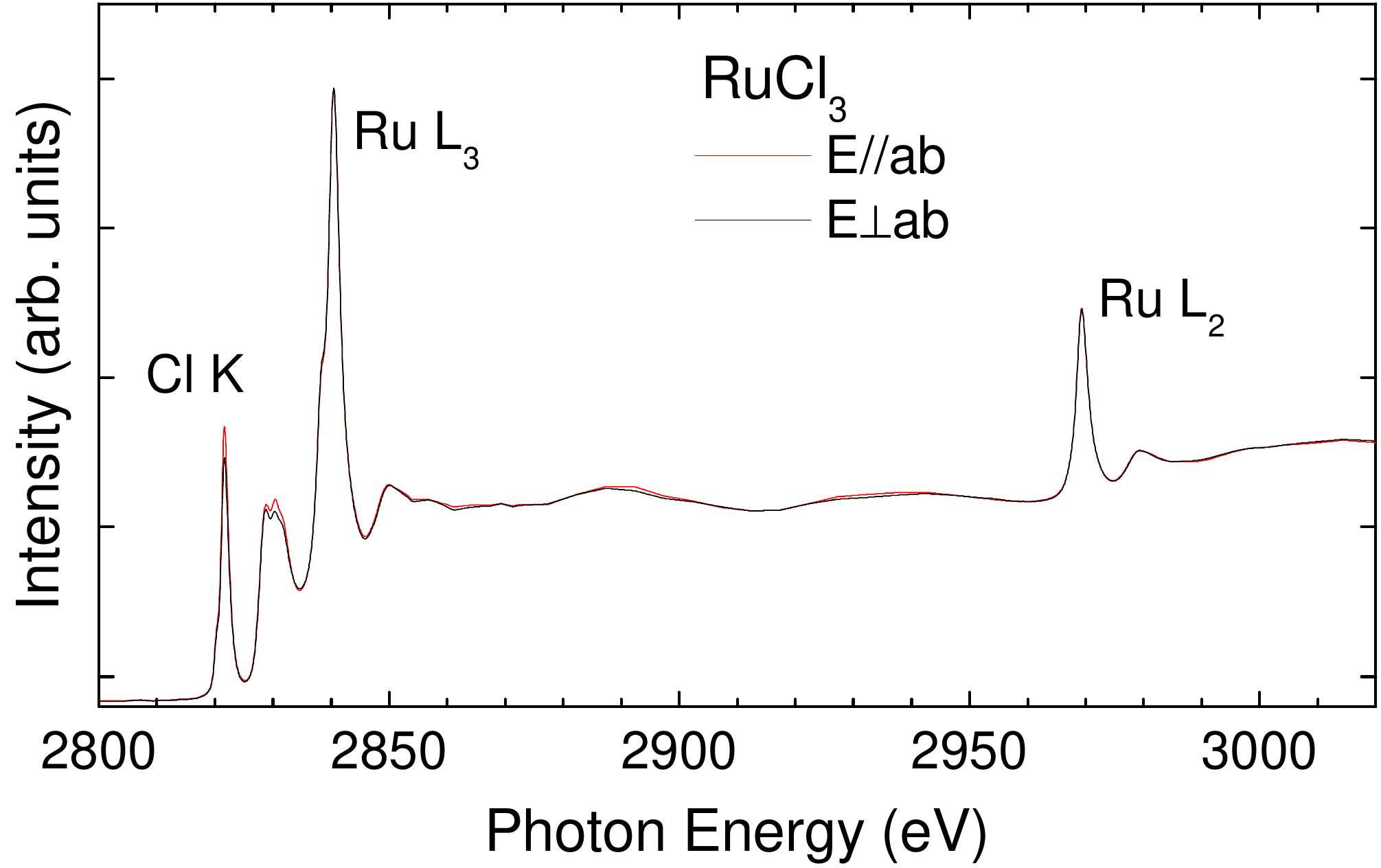}
\caption{(color online) Ru-$L_{2,3}$ XAS spectra of a RuCl$_3$ single crystal for incoming linear polarized light
with the electric field vector $\textbf{E}$ normal (black line) and
parallel (red line) to the $ab$ plane. }\vspace{-0.2cm}
\end{figure}

Starting from polycrystalline RuCl$_3$ (Chempur) the crystals were obtained by chemical
transport reaction with chlorine between 730 to 660 $^{\circ}$C. The crystals were
annealed for five months at 440 $^{\circ}$C under vacuum. A full
characterization of the crystals is provided in Ref.\,\onlinecite{Majumder15}.
The linear polarized XAS at the Ru-$L_{2,3}$ edges (2800-3000 eV) was measured at the
16A1 tender x-ray beamline of the NSRRC in Taiwan. The spectra were collected at room temperature
in the total electron yield (TEY) mode. The degree of linear polarization of the incident light was
close to 100\% and the energy resolution was set to 0.6~eV. The x-ray magnetic circular
dichroism (XMCD) experiments at the Ru-$L_{2,3}$ edges were performed at the BL29 Boreas
beamline of the ALBA synchrotron radiation facility in Barcelona. The energy resolution was
1.4~eV and the degree of circular polarization delivered by the Apple II-type elliptical
undulator was set to 70\%. The spectra were recorded in the TEY method
at $T$ = 2 K and $B$ = 6 T. The RuCl$_3$ crystals were cleaved $\emph{in situ}$
to obtain a clean sample surface normal to the (001) direction.
Density functional theory (DFT) based calculations were carried out using the full-potential
local-orbital code \texttt{FPLO} \cite{fplo1}, including both SOC and electron correlation
($U$) effects for the
simplest ferromagnetic spin configuration \cite{footnote1}.

 \begin{figure}[b]
\includegraphics[angle=0,width=\columnwidth]{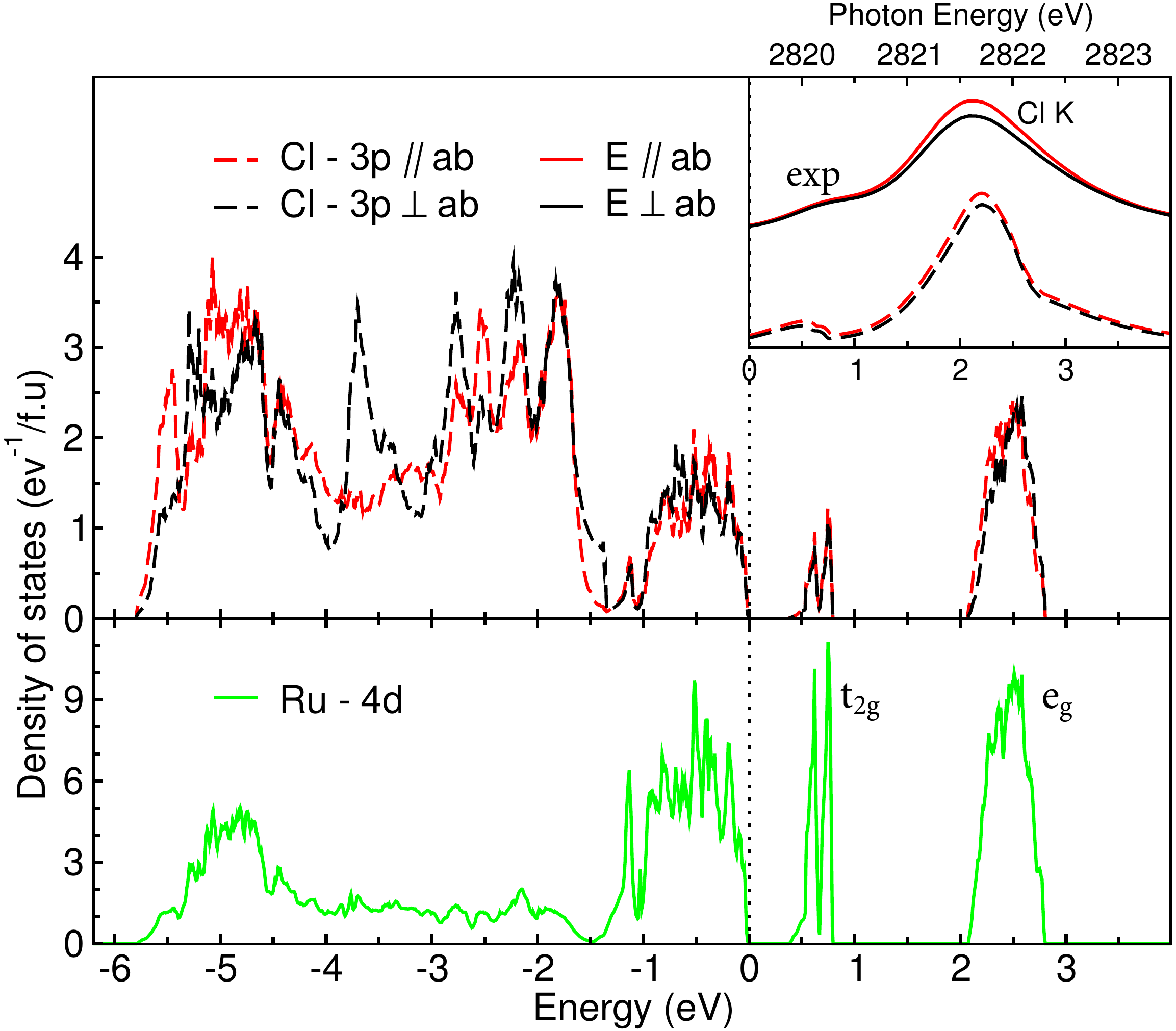}
\caption{(color online) Partial density of states of the Cl-$3p$ orbitals for different polarization (top) and
of the Ru-$4d$ orbitals (bottom). The inset compares the broadened unoccupied DOS of Cl-$3p$ to the Cl-$K$ edge of the measured XAS spectra.} \vspace{-0.2cm}
\end{figure}

In Fig.\,1 we report the Ru-$L_{2,3}$ XAS measured on RuCl$_3$ at room temperature for
linearly polarized light coming in with the electric field vector $\textbf{E}$ normal
and parallel to the $ab$-plane. The chosen geometry has the
incoming light polarization parallel and normal to the trigonal
axis [111] of the local $D_{3d}$ symmetry. The Ru $2p$ core-hole spin-orbit coupling splits the
spectrum roughly in two parts, namely the $L_{3}$ (at $h\nu \approx$~2840~eV) and the
$L_{2}$ (at $h\nu \approx$~2969~eV) white line regions. Additional features appearing
in the low energy part of the spectrum are related to the Cl-$K$ edge at $h\nu \approx$~2822~eV.

We first focus on the Cl-$K$ edge features, which can be explained in terms of
dipole allowed transitions from the Cl $1s$ core level to the unoccupied Cl $3p$ states.
Fig.\,2 displays the Cl $3p$ and Ru $3d$ partial density of states (DOS) from the DFT calculations, which reveal
the presence of two sharp features above the Fermi level, namely at $\sim$0.5 eV and $\sim$2.4 eV.
These are given by the unoccupied Ru $4d$ $t_{2g}$ and $e_g$ states, respectively,
hybridizing with the Cl $3p$. Comparing these unoccupied Cl states with the experimental Cl-$K$ edge
features, we can observe a very satisfactory agreement, especially when we include a broadening
for the calculated curves in order to take the experimental resolution into account. Also the weak
but clear polarization dependence in the experimental spectra is well explained by the DFT calculations.

For a better view of the multiplet and polarization dependence in the Ru-$L_{2,3}$ spectra
we show in Fig.\,3 a close-up of the spectra. Notably  there is hardly
any linear dichroism (LD) visible at the low energy peak (at 2838 eV) of the $L_{3}$ edge, which
corresponds to the signal of the $t_{2g}$ orbitals. The absence of LD is a very
sensitive signal for how close to cubic the local symmetry is. For example, a trigonal elongation
(compression) of the RuCl$_6$ octahedron will cause a splitting of the $t_{2g}$ orbitals in to
$a_{1g }$ and $e_g^{\pi}$ orbitals, with the $a_{1g }$ orbital lying higher (lower) in energy and,
hence, having more (less) holes. Such an uneven hole distribution will then produce a difference in
the spectral weight between $\textbf{E}$ normal and parallel to the [111] axis. The experimental result
that the LD is vanishingly small gives a clear and direct indication that the trigonal distortion of
the RuCl$_6$ octahedra is electronically negligible.

\begin{figure}[t]\centering
\includegraphics[width=\linewidth]{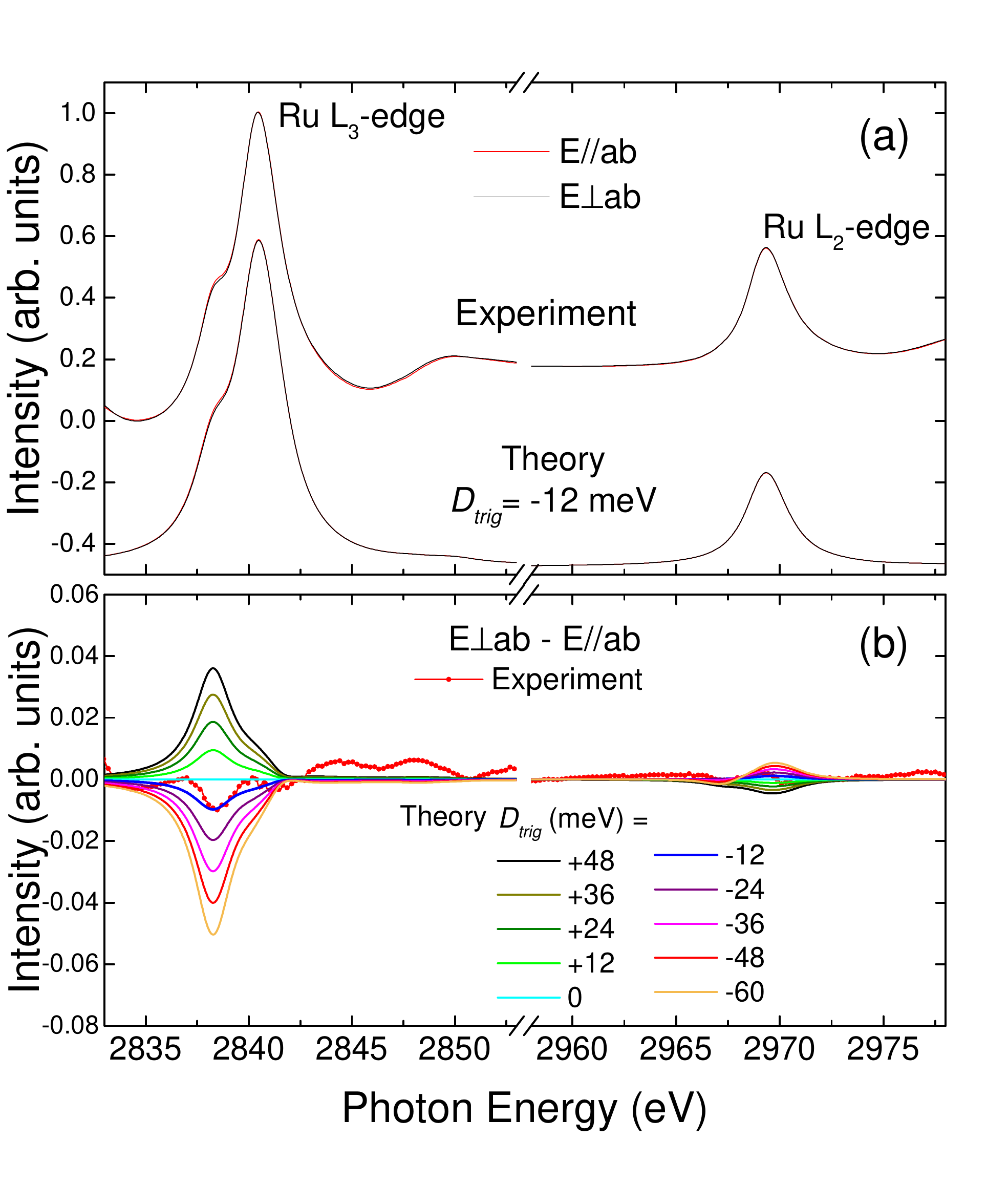}
\caption{(color online) (a): Close-up of the Ru-$L_{2,3}$ XAS spectra of a RuCl$_3$ single crystal for incoming linear polarized light with the electric field vector $\textbf{E}$ normal (black lines) and parallel
(red lines) to the $ab$ plane. (b): Experimental difference spectrum (red circles) compared with calculations for different values
of $D_{trig}$.}\vspace{-0.2cm}
\end{figure}

To obtain a quantitative estimate  of how close to cubic the system is from an electronic point
of view, we have simulated the Ru-$L_{2,3}$ XAS spectrum using the configuration-interaction cluster
model \cite{degroot94, thole97}. This model includes the full atomic multiplet theory and takes into
account the intra-atomic $4d-4d$ and $2p-4d$ Coulomb interactions, the atomic $2p$ and $4d$
spin-orbit couplings, the Cl-$3p$ with Ru-$4d$ hybridization, and the local crystal field parameters.
In the simulations we considered a RuCl$_6$ cluster with a D$_{3d}$ symmetry as further distortions
of the octahedra beyond the trigonal symmetry can be safely neglected \cite{Johnson15,Cao16}.
The cubic crystal field splitting between the Ru $t_{2g}$ and $e_g$ orbitals was estimated from the
difference in energy position between the maximum in the XAS spectrum, corresponding to the
signal from the unoccupied $e_g$ levels, and the maximum of the XMCD signal (see below), which
is due to the $t_{2g}$ orbitals. The hybridization parameters and the crystal field acting on the
chlorine ligands were extracted \textit{ab-initio} by DFT calculations.
The calculations of the XAS spectra were performed using
the XTLS 8.3 code \cite{Tanaka94} and the input parameters are given in Ref.\,\onlinecite{calc_par}.

The calculated Ru-$L_{2,3}$ XAS spectra are plotted in Fig. 3(a). They nicely reproduce the
experiment. In order to show how the trigonal distortion affects LD, we have plotted in
Fig. 3(b) the experimental difference spectrum ($\textbf{E}\bot ab - \textbf{E}//ab$)
together with the calculations for different trigonal crystal field splitting $D_{trig}=E(e_g^{\pi})-E(a_{1g})$.
As one can see, the LD is very sensitive to the magnitude and sign of the trigonal splitting.
For positive $D_{trig}$, the calculated LD has the opposite sign
compared to the experimental one. On the other hand, a negative
$D_{trig}= -24$ meV produces a LD signal with the correct sign but is already twice
as large compared to the experiment. Hence, our experimental LD signal provides strong
limits for the trigonal splitting of the $t_{2g}$ orbitals. The best fit to the experimental data
is obtained for $D_{trig}= -12$ meV. The accuracy of our method is actually limited by the
presence of the Cl $K$-edge EXAFS oscillations which occur in the same region as the
Ru-$L_{2,3}$ edges. Part of these EXAFS oscillations show up as a slow varying background
outside the Ru $L$-edge white line region (Fig. 3(b) ) and is as small as the small LD
in the Ru-$L_{2,3}$. Thus, our estimates result in a $D_{trig}= -12\pm10$ meV.
This is an important finding since we now can conclude that the trigonal crystal field splitting
is at least ten times smaller than the spin-orbit coupling constant ($\sim$150 meV), implying that
the Ru $d^5$ ion may indeed be in the $J_{\emph{eff}}=1/2$ ground state.

\begin{figure}[t]\centering
\includegraphics[width=\linewidth]{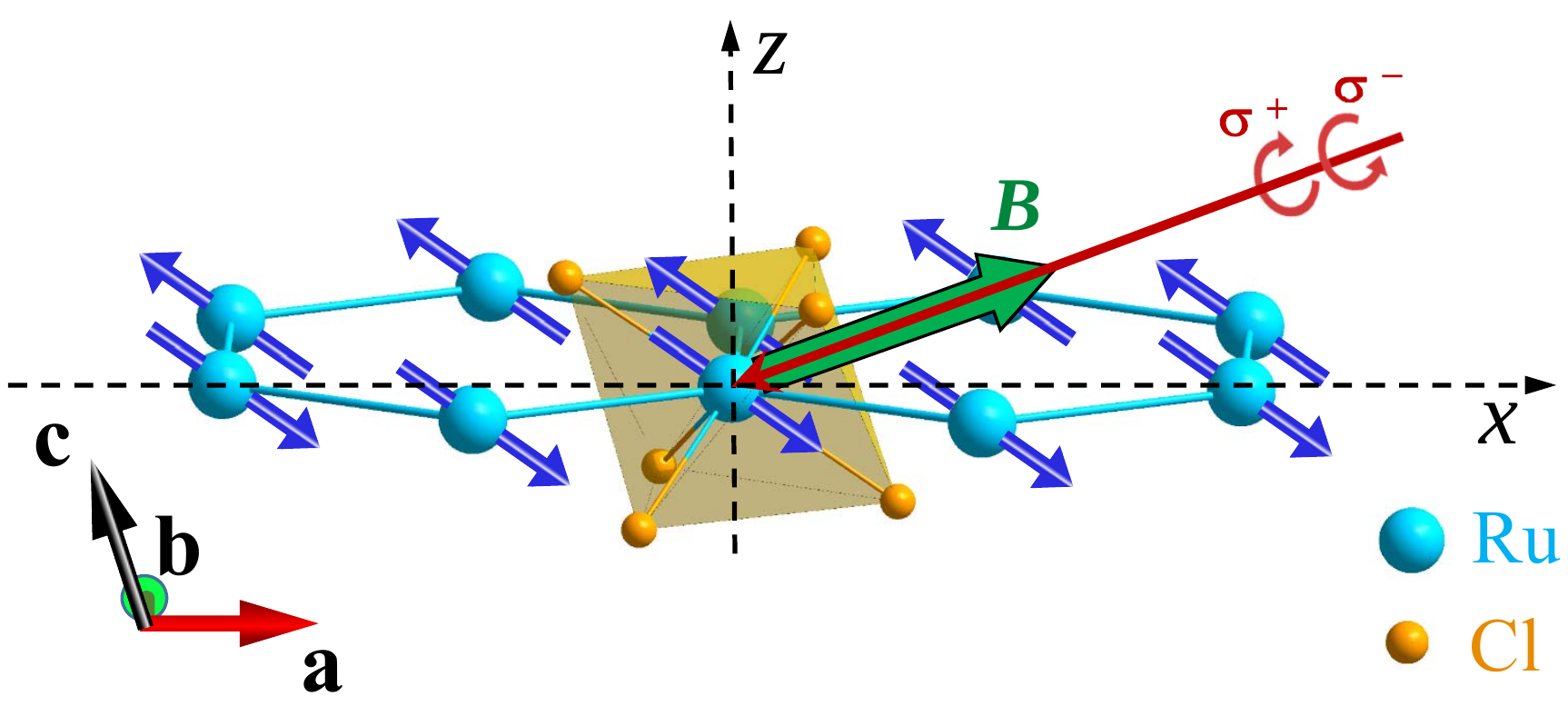}
\caption{(color online) Setup of the XMCD experiment: the magnetic field $B$ is applied parallel to the
Poynting vector of the circularly polarized photons and forms an angle of $20^{\circ}$ with
the $ab$ plane. }\vspace{-0.2cm}
\end{figure}

Having established the crystal field situation, we now investigate the magnetic ground state
of the Ru ions by performing Ru-$L_{2,3}$ x-ray absorption measurements using
circular polarized light with the photon spin aligned parallel ($\sigma^+$) and antiparallel ($\sigma^-$)
to the magnetic field. A sketch of the experimental geometry is shown in Fig. 4.
The difference or XMCD spectrum ($\sigma^+-\sigma^+$) and the sum spectrum
($\sigma^++\sigma^+$) are reported in Fig. 5. The spectra were collected at
$T$~=~2~K in grazing incidence with the magnetic field ($B$~=~6~T) lying in the $ac$ plane
($xz$ plane in local D$_{3d}$ symmetry) and forming an angle of $20^{\circ}$ with the (100)
axis ($x$-axis in local symmetry). The grazing geometry allowed to maximize the magnetic
signal according to the easy-plane magnetic anisotropy of RuCl$_3$ reported in literature \cite{Kubota15,Majumder15}.

\begin{figure}[t]\centering
\includegraphics[width=\linewidth]{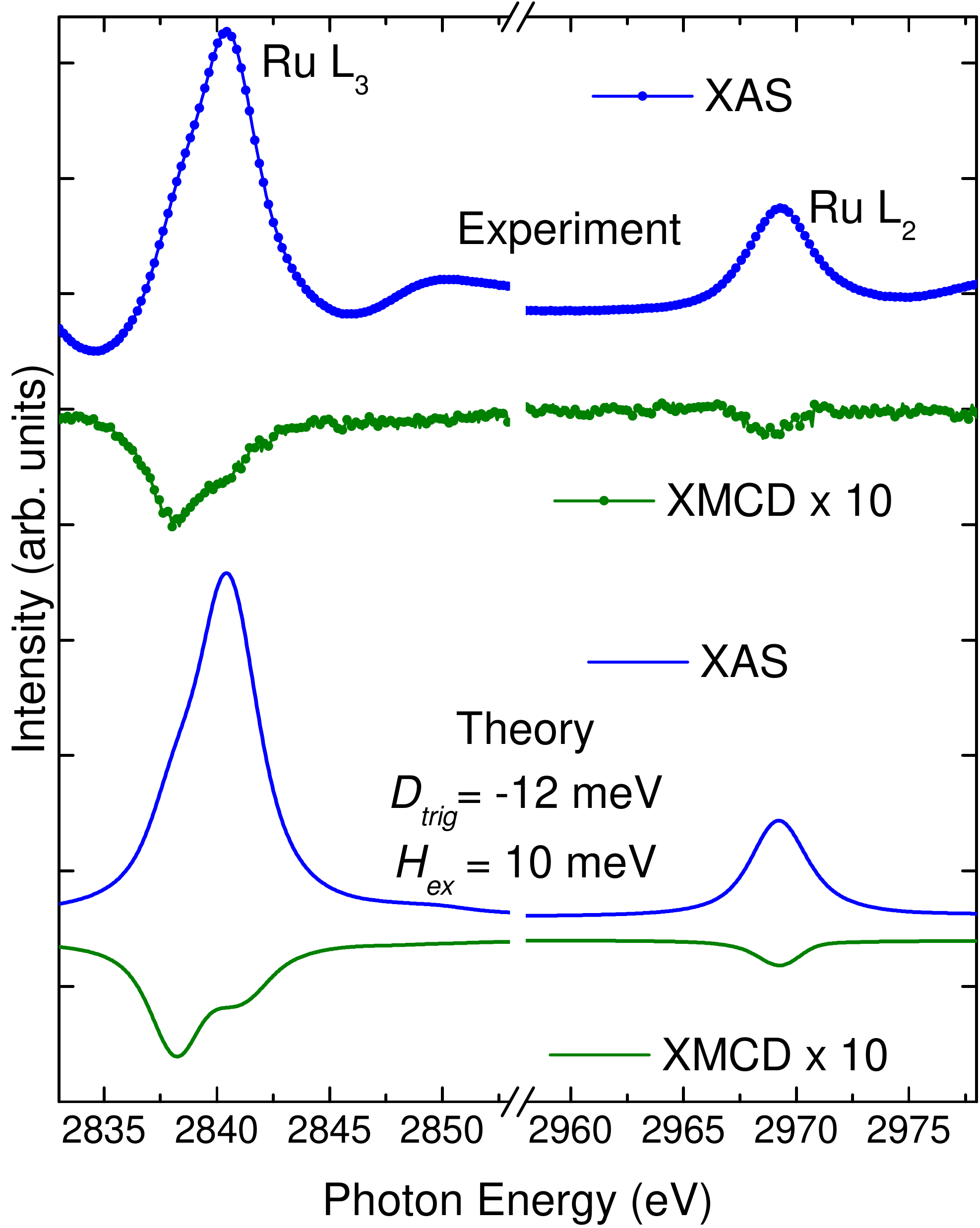}
\caption{(color online) Experimental Ru-$L_{2,3}$ XAS (blue circles) and XMCD (green circles) spectra of a
RuCl$_3$ single crystal together with calculated XAS (blue line) and XMCD (green line). The
spectra were measured at $T$ = 2 K and $B$ = 6 T.}\vspace{-0.2cm}
\end{figure}

The Ru-$L_{2,3}$ XMCD spectrum as obtained from our full-multiplet calculations is also presented
in Fig. 5. In our model we have used the same parameters as for the simulation of the LD data,
with $D_{trig}=-12$~meV. The lineshape of the calculated XMCD spectrum
is in nice agreement with the experimental one, further validating our calculations. Here, we have
used an exchange field of about $H_{ex}=10$~meV in order to reproduce the magnitude of the
experimental XMCD spectrum. If we would have used zero exchange field, the calculated XMCD
signal in an applied field of 6 Tesla is much larger than the measured one, i.e. the magnitude of
the XMCD is very sensitive to the size of the exchange field (see supplemental
information). This in fact can be understood as RuCl$_3$ exhibits a zig-zag
modulated antiferromagnetic order below $T_N$~=~8~K and the XMCD signal is given only
by the canting of the moments induced by the applied field. The exchange field we have applied
is directed along the Ru spins, which form an angle of $\phi=35^{\circ}$ with the $ab$ plane
\cite{Cao16} (blue arrows in Fig. 3). We would like to note that the XMCD alone is rather unsuitable
to determine accurately the magnitude of the trigonal crystal field splitting (see supplemental information).

Having obtained $D_{trig}$ from LD and $H_{ex}$ from XMCD we can now focus on the
orbital moment of Ru in RuCl$_3$. Very interestingly, the XMCD signal has the same negative
sign at both $L_3$ and $L_2$ edges. Usually, the XMCD has the opposite sign at the
two edges, which is a consequence of the reduction of the orbital moment from its atomic value
when the transition metal ion is placed in a solid. The fact that the XMCD does not change sign
clearly indicates that the orbital moment of the Ru$^{3+}$ ion in RuCl$_3$ is large, possibly
close to the atomic values. From our configuration-interaction calculations we obtain a ratio of
$L_x/2S_x=2.0$ for $H_{ex}=10$ meV (2.1 if $H_{ex}=0$ meV). This value is very close to the
ratio between the orbital and the spin moments expected for a pure $J_{\emph{eff}}=1/2$ system
\cite{Kim2008}.

From the Zeeman splitting of the energy levels in the presence of an applied magnetic field we
can calculate the magnetic $g$ factor to be $g_x=g_y=2.27$ and $g_z=2.05$ \cite{g_factor}.
The fairly isotropic g factor ($g_x/g_z=1.1$) indicates that the strong anisotropy shown by both
susceptibility \cite{Kubota15,Majumder15} and high-field magnetization \cite{Kubota15,Johnson15}
can not be ascribed to single ion physics. Instead, hybridization with neighboring Ru ions needs
to be considered explicitly, giving rise to various nearest neighbor and next nearest neighbor
Heisenberg, Kitaev and off-diagonal couplings \cite{Winter2016}.
Our calculations also shows that the average $g=(2g_x+g_z)/3=2.2$
is larger than that (2.0) for a pure ionic $t_{2g}^5$ system. While covalency tends to
decrease the value of the $g$ factor, the mixing-in of some $e_g$ character into the $t_{2g}$
manifold will quickly increase the $g$ factor value. This $e_g$-$t_{2g}$ mixing can take place
locally on a one-electron level, for example by the presence of a trigonal crystal field, but also
(and in fact, certainly) on a many-electron level due to the presence of atomic multiplet
interactions (Slater $F^2$ and $F^4$ integrals) which are not at all small compared to the
$e_g$-$t_{2g}$ crystal field splitting ($10Dq$).

To summarize, our X-ray absorption linear dichroism study demonstrates that the ground state
of RuCl$_3$ is a doublet with a very close to perfect cubic local symmetry. Our excellent simulations
of the experimental spectra give a ratio of 2.0 between the orbital and the spin contributions to the
local Ru $4d$ magnetic moment, i.e. the value expected for a $J_{\emph{eff}}=1/2$ ground state. Further quantitative modeling is highly desired as to include not only the Ru $t_{2g}$ but also the $e_g$ orbitals.

\begin{acknowledgments}
We would like to thank the NSRRC and ALBA for providing us with beam time and for the
support from the staff during the experiments. The research in Dresden was partially supported
by the Deutsche Forschungsgemeinschaft through SFB 1143 and FOR1346. K.-T. Ko acknowledges
support from the Max Planck-POSTECH Center for Complex Phase Materials (No. KR2011-0031558).
\end{acknowledgments}

\clearpage
\newpage
\newpage

%% file: RuCl3_supp.tex
\title{Electronically highly cubic conditions for Ru in $\alpha$-RuCl$_3$}
\author{S.~Agrestini}
  \affiliation{Max Planck Institute for Chemical Physics of Solids,
     N\"othnitzerstr. 40, 01187 Dresden, Germany}
\author{C.-Y.~Kuo}
  \affiliation{Max Planck Institute for Chemical Physics of Solids,
     N\"othnitzerstr. 40, 01187 Dresden, Germany}
\author{K.-T.~Ko}
  \affiliation{Max Planck Institute for Chemical Physics of Solids,
     N\"othnitzerstr. 40, 01187 Dresden, Germany}
\author{Z.~Hu}
  \affiliation{Max Planck Institute for Chemical Physics of Solids,
     N\"othnitzerstr. 40, 01187 Dresden, Germany}
\author{D.~Kasinathan}
  \affiliation{Max Planck Institute for Chemical Physics of Solids,
     N\"othnitzerstr. 40, 01187 Dresden, Germany}
\author{H.~Babu~Vasili}
  \affiliation{ALBA Synchrotron Light Source, E-08290 Cerdanyola del Vall\`{e}s, Barcelona, Spain}
\author{J.~Herrero-Martin}
  \affiliation{ALBA Synchrotron Light Source, E-08290 Cerdanyola del Vall\`{e}s, Barcelona, Spain}
\author{S.~M.~Valvidares}
  \affiliation{ALBA Synchrotron Light Source, E-08290 Cerdanyola del Vall\`{e}s, Barcelona, Spain}
\author{E.~Pellegrin}
  \affiliation{ALBA Synchrotron Light Source, E-08290 Cerdanyola del Vall\`{e}s, Barcelona, Spain}
\author{L.-Y. Jang}
  \affiliation{National Synchrotron Radiation Research Center, 101 Hsin-Ann Road, Hsinchu 30076, Taiwan}
\author{A. Henschel}
 \affiliation{Max Planck Institute for Chemical Physics of Solids,
     N\"othnitzerstr. 40, 01187 Dresden, Germany}
\author{M. Schmidt}
 \affiliation{Max Planck Institute for Chemical Physics of Solids,
     N\"othnitzerstr. 40, 01187 Dresden, Germany}
\author{A.~Tanaka}
  \affiliation{Department of Quantum Matter, ADSM, Hiroshima University, Higashi-Hiroshima 739-8530, Japan}
\author{L.~H.~Tjeng}
  \affiliation{Max Planck Institute for Chemical Physics of Solids,
     N\"othnitzerstr. 40, 01187 Dresden, Germany}

\date{\today}

\maketitle

\begin{figure}
\includegraphics[width=1.0\linewidth]{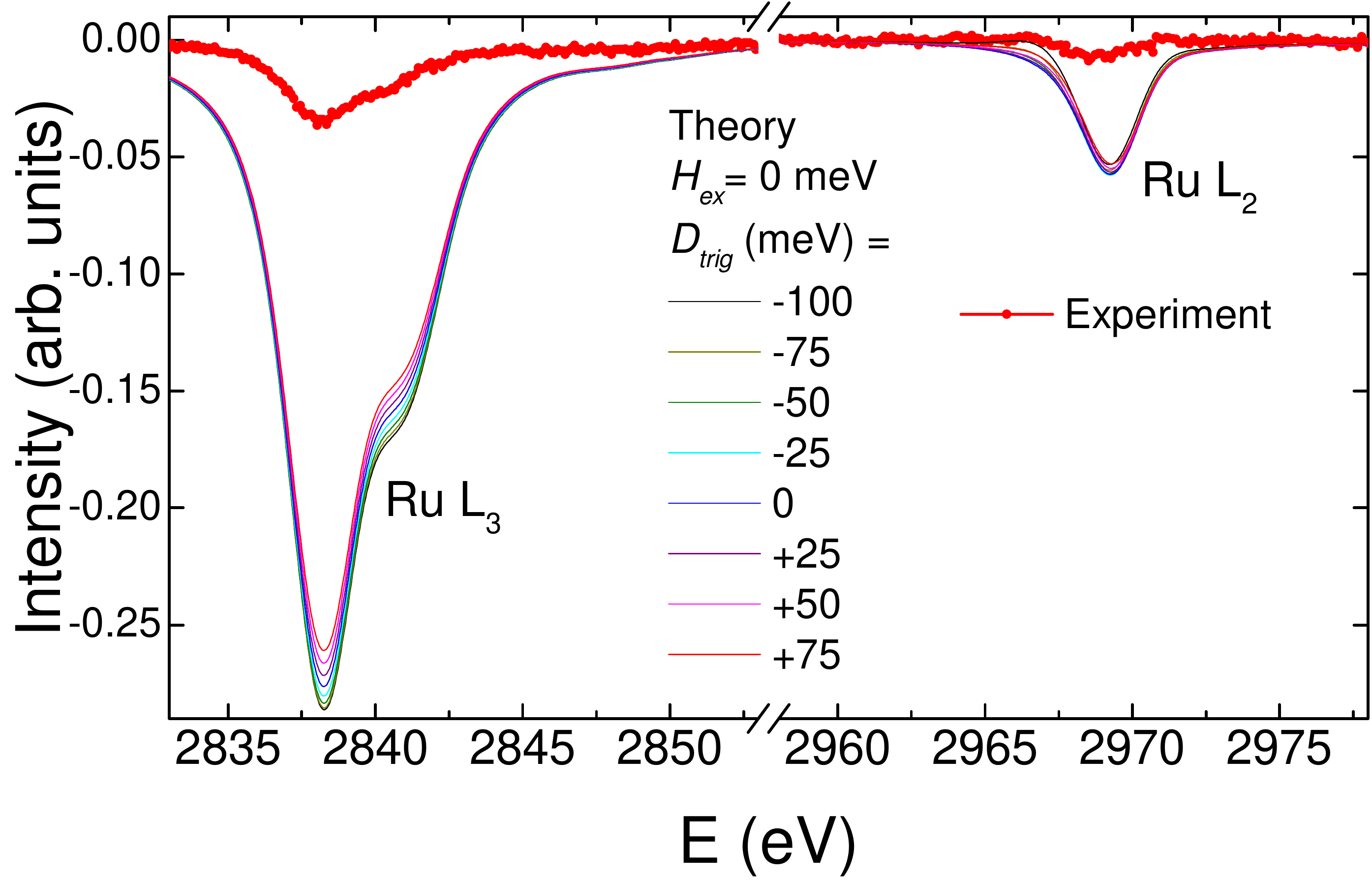}
\justify
FIG. S1: (color online) Ru-$L_{2,3}$ XMCD simulations (solid lines) calculated using different values of the trigonal crystal
field splitting $D_{trig}$ with the exchange field $H_{ex}$ set to zero. The calculations were done for $T$ = 2 K and
$B$ = 6 T, the conditions at which the experimental XMCD spectrum of RuCl$_3$ was taken (red circles).
The intensity was normalized to have the calculated Ru-$L_{3}$ XAS peak height to match the experimental one.
\vspace{-0.2cm}
\end{figure}

In Fig. S1 we report simulations for the Ru-$L_{2,3}$ XCMD spectra calculated for $T$ = 2 K and $B$ = 6 T using
different values of the trigonal crystal field splitting $D_{trig}$ under the assumption that the exchange field $H_{ex}$
is absent. One can clearly observe that the simulated XMCD signal is always much larger than the measured one of
RuCl$_3$, no matter what trigonal distortion is considered.

\begin{figure}[ht]\centering
\includegraphics[width=\linewidth]{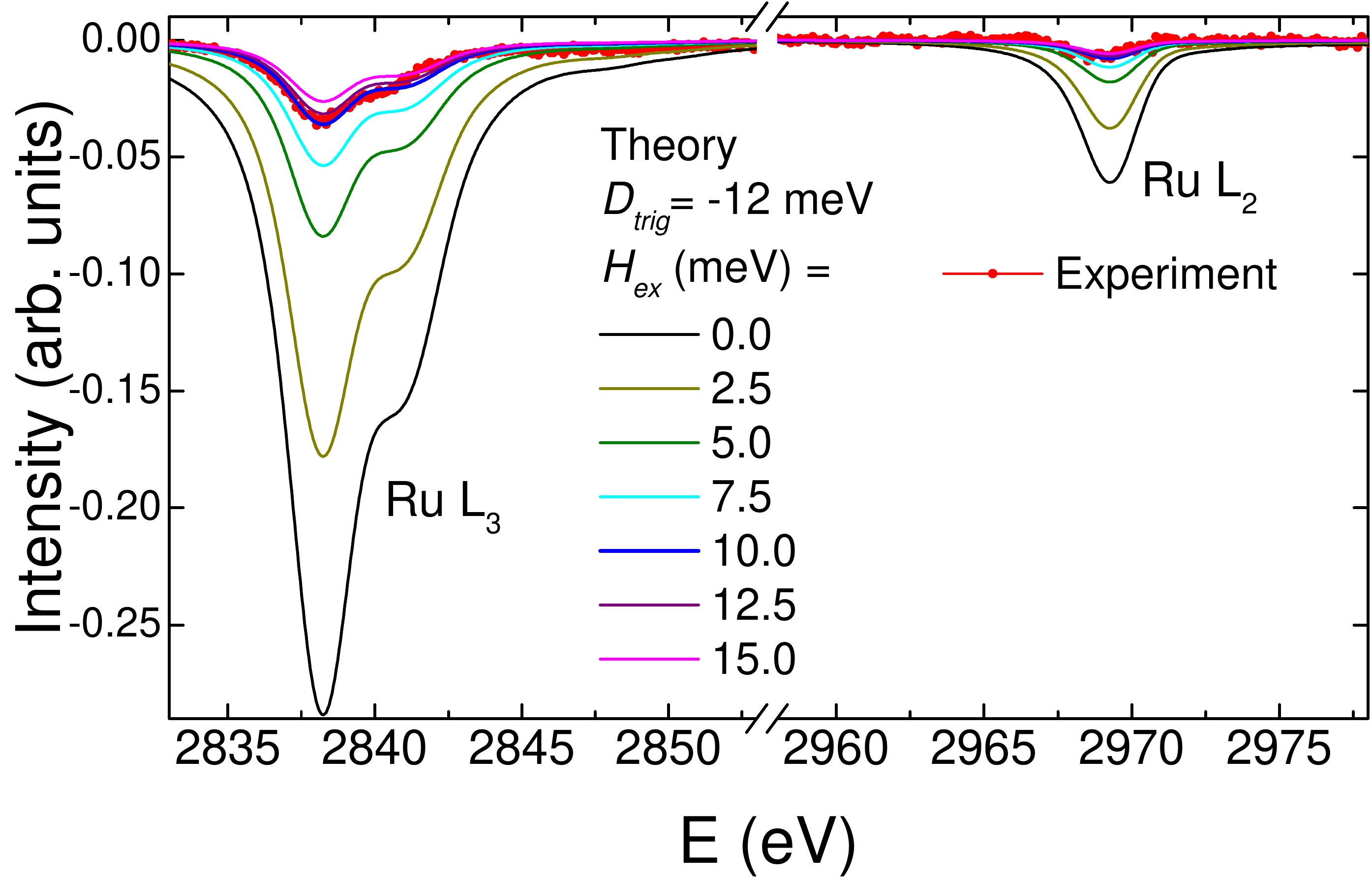}
\justify
FIG. S2: (color online) Ru-$L_{2,3}$ XMCD simulations (solid lines) calculated using different values of $H_{ex}$ with
$D_{trig}$ set at -12 meV.  The experimental XMCD spectrum of RuCl$_3$ is also given (red circles). The intensity
was normalized to have the calculated Ru-$L_{3}$ XAS peak height to match the experimental one.
\vspace{-0.2cm}
\end{figure}

In Fig. S2 we show the simulations for the XMCD for different values of $H_{ex}$ with $D_{trig}$ fixed at -12 meV.
The exchange field has a very strong influence on the magnitude of the XMCD signal. The experimental XMCD is
best simulated for $H_{ex}$ = 10 meV. We note that in the considered range, the exchange field has no significant
effect on the lineshape of the XMCD spectrum but determines mainly the size of the XMCD signal.

\begin{figure}[t]
\includegraphics[width=1.0\linewidth]{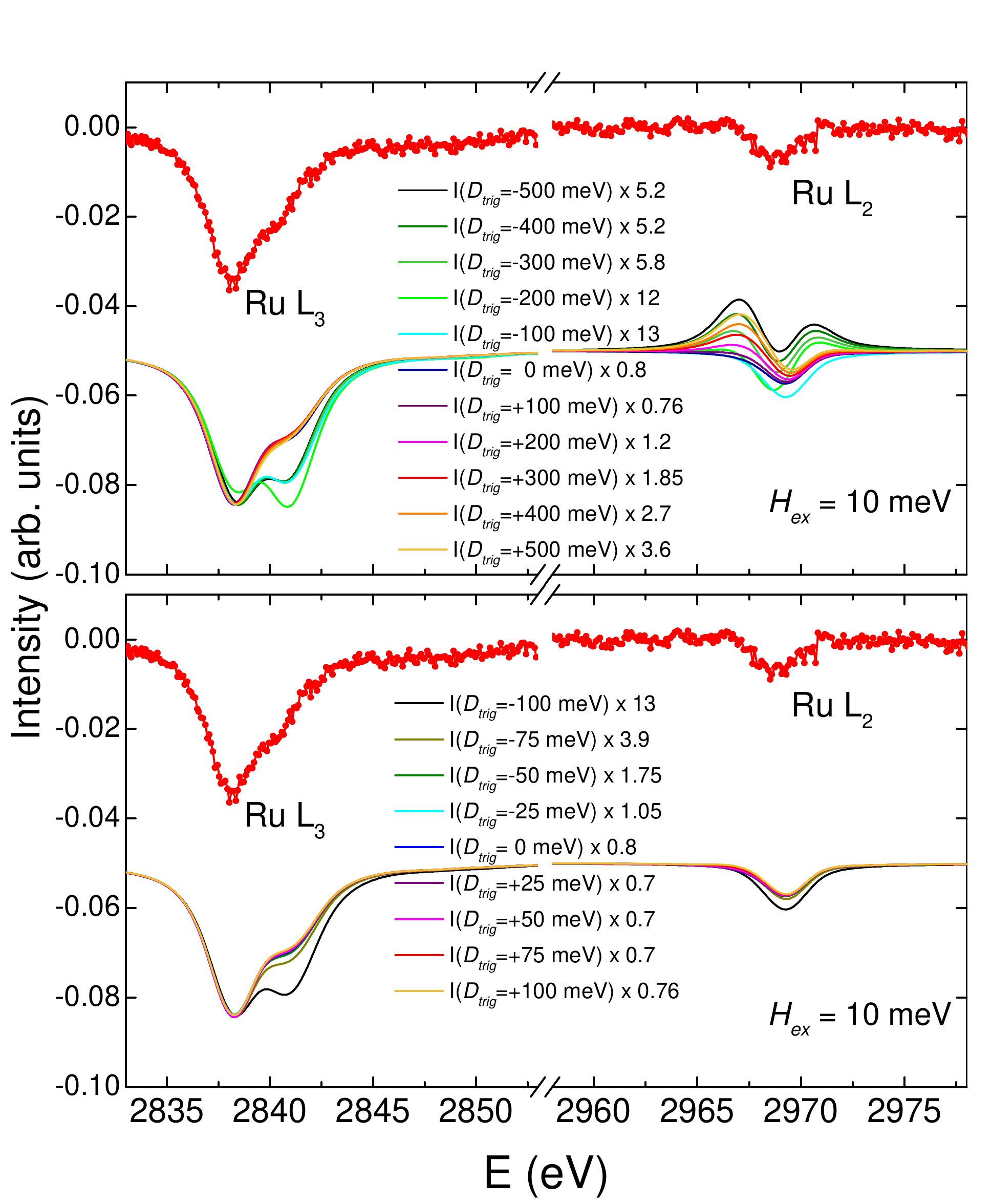}
\justify
FIG. S3: (color online) Ru-$L_{2,3}$ XMCD simulations (solid lines) calculated using different values of $D_{trig}$ with
$H_{ex}$ = 10 meV, together with the experimental XMCD spectrum of RuCl$_3$ (red circles). In the top
panel $D_{trig}$ is varied between -500 and +500 meV, and in the bottom panel between -100 and +100
meV. For a better comparison of the line shapes, the simulated spectra are normalized to the XMCD peak
height, with the normalization factor also indicated in the legend.
\vspace{-0.2cm}
\end{figure}

Fig. S3 displays the simulations for the XMCD as a function of $D_{trig}$ in order to investigate the
sensitivity of the XMCD lineshape to $D_{trig}$. The calculations have been done for a fixed $H_{ex}$ = 10 meV
and the curves have been normalized to the XMCD peak height to highlight the lineshape of the spectra
rather than the magnitude of the signal. The top panel shows the results for a large range of $D_{trig}$,
i.e. between -500 and +500 meV, and the bottom panel presents the result for a narrower range with
smaller steps, i.e. between -100 and +100 meV. From the figure we can observe that the spectral lineshape of
XMCD is sensitive to the trigonal crystal field splitting, but only so if the $D_{trig}$ value is varied in a range
that exceeds the spin-orbit coupling constant of about 150 meV. Varying $D_{trig}$  between -50 and +100 meV
in fact does not change the lineshape visibly. In other words, XMCD alone is not a very sensitive method to
determine $D_{trig}$ accurately. To establish whether the $J_{\emph{eff}}=1/2$ condition is fulfilled (or not) implies
that one needs to determine whether $D_{trig}$ is negligible (or not negligible) compared to the spin-orbit
constant. The more accurate method to resolve this issue is to measure the smallness of the linear dichroism
in the XAS itself. XMCD is valuable to determine the magnitude of the exchange field and to serve as a strict
consistency check of the simulations: the same set of parameters must provide excellent simulations for both
the XAS and the XMCD spectra.

In literature the sum rules for XMCD developed by Thole and Carra et al. \cite{thole92a,Carra93} are
often used to extract directly from the spectrum the ratio of orbital and spin moments. The sum rules
can be summarized as:

\begin{equation}\label{eq:ratio}
\frac{L_x}{2S_x+7T_x}=\frac{2}{3}\cdot \frac{\int_{L_{2,3}}(\sigma^+-\sigma^-)dE}{\int_{L_{3}}(\sigma^+-\sigma^-)dE-2\int_{L_{2}}(\sigma^+-\sigma^-)dE}.
\end{equation}

where $S_x$ and $L_x$ are the spin and orbital contributions to the local magnetic moment,
respectively, and $T_x$ is the intra-atomic magnetic dipole moment. The application of the sum rules
to our experimental Ru $L_{2,3}$ XMCD data gives $L_x/(2S_x+7T_x)=1.0(1)$, which confirms that
the orbital moment of Ru$^{3+}$ is relatively large. If we use the value of $T_x/2S_x=0.15$ given
by our configuration-interaction calculations, we obtain from the application of the sum rules the ratio $L_x/2S_x=2.0(1)$, which is in very good agreement with the value given by our configuration
interaction calculations for $D_{trig}=5$~meV and $H_{ex}=10$ meV. This demonstrates that the
analysis using the sum rules is fully consistent with the analysis based on the simulations of the
XAS and XMCD spectra, i.e. analyses based on the lineshape, as it should be.

\begin{figure}[t!]
\includegraphics[width=1.0\linewidth]{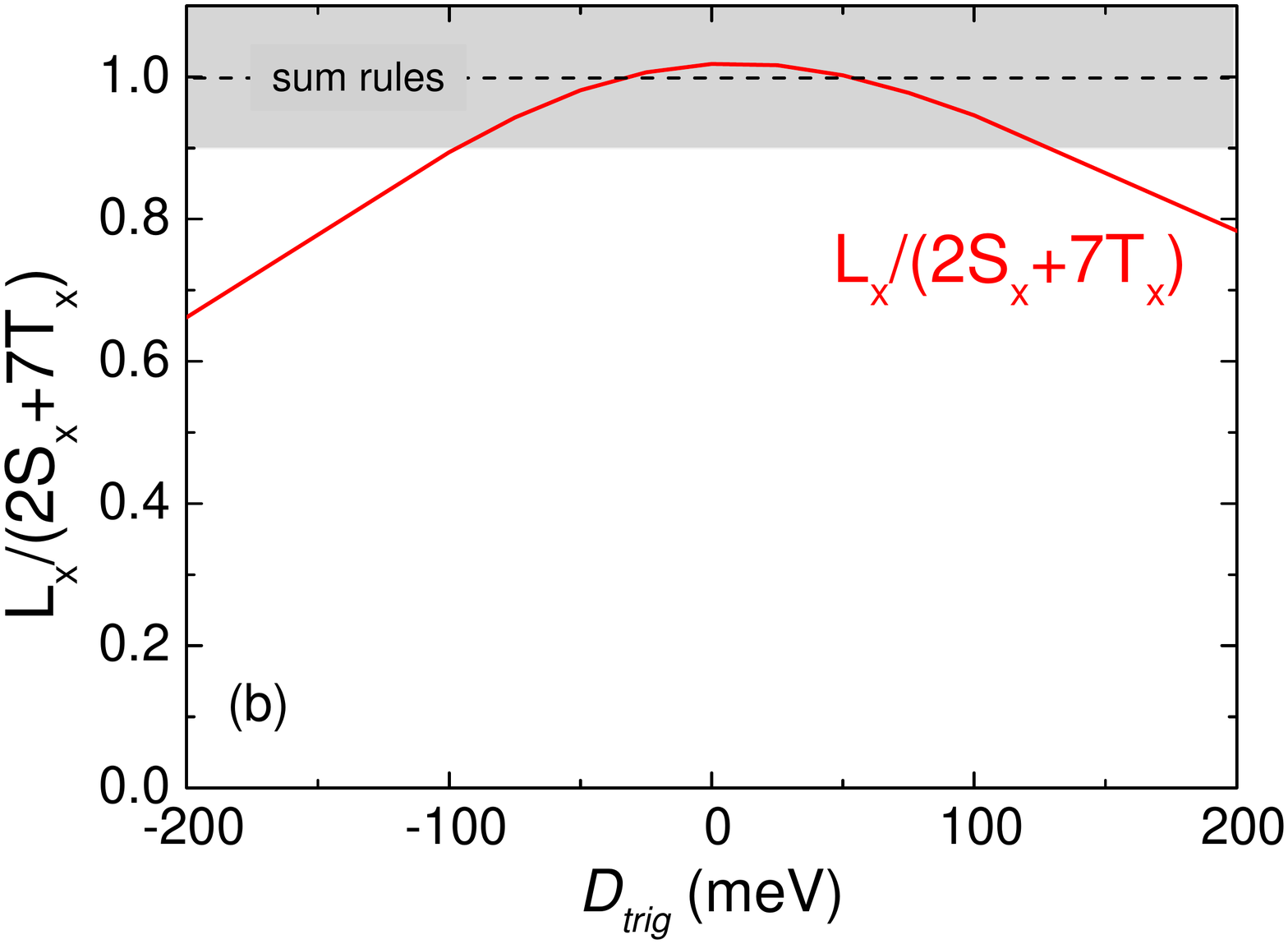}
\justify
FIG. S4: (color online) Calculated $L_x/(2S_x+7T_x)$ for $H_{ex}=10$ meV as a function of $D_{trig}$. The
horizontal dashed line indicates the $L_x/(2S_x+7T_x)$ value given by the sum rules applied to
the experimental XMCD data and the shadow area represents its error bar.
\vspace{-0.2cm}
\end{figure}

Fig. S4 shows the calculated $L_x/(2S_x+7T_x)$ ratio as a function of the trigonal crystal field splitting
for a fixed exchange field of 10 meV. We can clearly observe that the sum rule quantity $L_x/(2S_x+7T_x)$
is a slowly varying function of $D_{trig}$. The experimental XMCD data is given by the dashed line
and the shadow area represents its error bar. Hence, the value obtained by applying the sum rules to the
XMCD data can correspond to the wide range $-100\leq D_{trig}\leq +100$~meV. In other words, our
calculations reveal that the XMCD alone of RuCl$_3$ cannot, either through spectral lineshape analysis or
through the application of the sum rules, provide a value of $D_{trig}$ with the required precision.
A reliable and accurate determination of $D_{trig}$ can be best obtained from the linear dichroic
signal in the polarization dependent XAS spectra as we have reported in the main text of the manuscript.

We would like to note that in our model we have taken into account the zig-zag modulated magnetic
structure (shown in fig. 3) proposed by Cao et al. \cite{Cao16}, by applying an exchange field of
10~meV along the direction of the Ru spins, which forms an angle of $\phi=35^{\circ}$ with the
$ab$ plane. If the alternative $\phi=-35^{\circ}$ zig-zag magnetic structure is considered then
the exchange field has to be reduced to 3~meV in order to simulate the size of the experimental
XMCD signal. The reason of the need of reducing the exchange field lies in the fact that in the
case of the second magnetic structure the applied magnetic field would be close to the direction
of the spins, which is a hard direction for the magnetization. With the present XCMD data we are
not able to distinguish which of the two magnetic structure is correct. Yet, this does not affect
the conclusions that we have made about the trigonal crystal field value and the $L_x/2S_x=2.0$
for RuCl$_3$, i.e. our finding that the Ru ground state fulfills the $J_{\emph{eff}}=1/2$ condition is robust.


%% file: RuCl3_combined.bbl
\begin{thebibliography}{60}
\bibitem{footnote1}The Coulomb repulsion $U$ was set to 1.5 eV and the Hund's exchange to 0.3 eV, following Refs.\,\onlinecite{Plumb14, Majumder15}.
\bibitem{calc_par} $U_{dd}$ = 4.0 eV, $U_{pd}$ = 5.5 eV, charge transfer energy $\Delta_{CT}=5.0~$eV,
    SOC~=~0.15~eV, ionic crystal field $10Dq^{ion}=1.5$~eV, $D_{trig}^{ion}=-12.0$~meV, hybridization
    $V(e_g^{\sigma})=3.336$~eV, $V(e_g^{\pi})=V(a_{1g})=1.76$~eV, ligand crystal field $10Dq^{lig}=1.3$~eV.
    Slater integrals were reduced to 70~\% of the Hartree-Fock values. The effective cubic crystal field is
    $10Dq^{eff}=2.35$~eV.  Gaussian (0.4 eV for the XAS in Fig. 3 and 1.4 eV for the XAS in Fig. 5) and
    Lorentzian (2 eV) broadenings were applied in order to take into account experimental resolution and
    lifetime effects, respectively.
\bibitem{g_factor} In our estimation of the $g$ factor we considered $H_{ex}=0$~meV.
\end{thebibliography}
